\documentclass{article}
\usepackage{spconf,amsmath,amsfonts,graphicx}
\usepackage{subcaption} 
\usepackage{url}
\usepackage{xcolor}

\title{Deep Snapshot HDR Reconstruction Based On the Polarization Camera}

\name{Juiwen Ting$^{\star}$ \qquad Xuesong Wu$^{\dagger}$ \qquad Kangkang Hu$^{\ddagger}$ \qquad Hong Zhang$^{\star}$}
  
\address{$^{\star}$ Department of Computing Science, University of Alberta, Canada \\
$^{\dagger}$ College of Intelligence Science and Technology, NUDT, China\\
$^{\ddagger}$ Huawei Fields Lab, China}
      
\begin{document}
\setlength{\belowdisplayskip}{4pt} \setlength{\belowdisplayshortskip}{4pt}
\setlength{\abovedisplayskip}{4pt} \setlength{\abovedisplayshortskip}{4pt}

\maketitle
\begin{abstract}
The recent development of the on-chip micro-polarizer technology has made it possible to acquire four spatially aligned and temporally synchronized polarization images with the same ease of operation as a conventional camera. In this paper, we investigate the use of this sensor technology in high-dynamic-range (HDR) imaging. Specifically, observing that natural light can be attenuated differently by varying the orientation of the polarization filter, we treat the multiple images captured by the polarization camera as a set captured under different exposure times. In our approach, we first study the relationship among polarizer orientation, degree and angle of polarization of light to the exposure time of a pixel in the polarization image. Subsequently, we propose a deep snapshot HDR reconstruction framework to recover an HDR image using the polarization images. A polarized HDR dataset is created to train and evaluate our approach. We demonstrate that our approach performs favorably against state-of-the-art HDR reconstruction algorithms.
\end{abstract}


\section{Introduction} \label{sec:intro}
The dynamic range of real-world scenes varies over several orders of magnitude. However, standard digital camera sensors can only capture a limited fraction of this range. The resulting low-dynamic-range (LDR) images often have over- or under-exposed areas that cannot reflect our visual ability to perceive details in both bright and dark regions of a scene. Furthermore, information is lost in these regions which make vision tasks such as object detection \cite{Lin_objectDetect} and visual tracking \cite{Song_visualTrack} difficult. High-dynamic-range (HDR) imaging seeks to recover the lost information, and aims to produce an image that captures a broader range of illumination than is available from the sensor directly.

Classical HDR imaging methods merge a sequence of LDR images at different exposures to create an HDR image \cite{Debevec, Mann}. However, this approach could create ghosting artefacts when there is motion between the input images. Although some methods can compensate for image motion \cite{Yan_AttnHDR, Lee_ExpBlendHDR}, these approaches still require multiexposure images as input, which may not always be available and practical. One-shot HDR techniques overcome these limitations by reconstructing an HDR image using one input image. As a result, they have attracted considerable attention in recent years. 

Several one-shot HDR works proposed to recover the brightness of saturated areas by extrapolating the light intensity using heuristic rules \cite{Akyuz, Masia}. However, these methods rely on hand-crafted features, and assumptions about the scene that might not adequately represent the real luminance value for proper pixel recovery. Recently, deep neural networks (DNNs) have improved HDR reconstruction results \cite{Eilertsen, Marnerides, Liu, Endo}. As another one-shot approach, DNNs based on a snapshot HDR sensor have also been studied \cite{Nayar, Suda}. In \cite{Suda}, for example, a DNN using multiexposure color filter array images is proposed for HDR imaging.   

In this paper, we present our method on using the polarization camera for deep snapshot HDR reconstruction. We design our approach based on two main observations. First, when using images captured by standard digital camera sensors, utilized by most existing approaches, the network is unable to hallucinate details for images with wide saturated areas. Second, information in a single polarization image is often insufficient, and this leads to poor results in saturated areas. To address these limitations, we propose to use the combination of polarization data and a fused input for the DNN to recover details lost in the saturated areas, in addition to expanding the dynamic range of unsaturated regions. The fusion mechanism is performed by a pixel-weighting function that combines the polarization images to reveal more details. Our contributions can be summarized as follows:

\begin{figure*}[t!]
\centering
\includegraphics[width=\textwidth,height=10em]{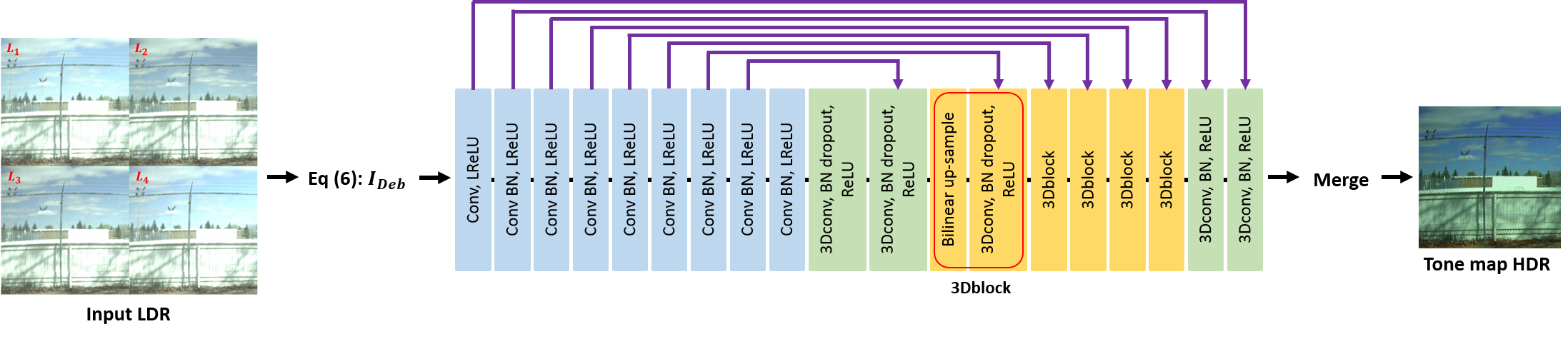}
\caption{Overview of our deep snapshot HDR reconstruction framework. It first estimates HDR luminance $I_{Deb}$, and then predicts the multiexposure images, followed by a merge and tone mapping step to output an HDR image. The autoencoder structure follows \cite{Endo}, where we replace the deconvolution operation with a bilinear interpolation operation.}
\label{fig:PFHDRNet}
\vspace{-0.5\baselineskip}
\end{figure*}

\begin{itemize}
\itemsep0em 
\vspace{-0.2cm}
\item We propose a novel deep learning framework to solve the snapshot HDR reconstruction problem using polarization images.
\vspace{-0.1cm}
\item We introduce a fusion mechanism based on the domain knowledge of polarization in combining images as input to the network.
\vspace{-0.1cm}
\item We collect a polarized HDR dataset for training and evaluation, and such a dataset of polarization images is currently unavailable in the literature.
\end{itemize}

\section{Method} \label{sec:method}
In this section, we first review the polarization image formation model, motivate why this image modality helps HDR reconstruction \cite{xuesong}, and then describe our framework for including the cues in the deep learning model.
\vspace{-0.37cm}
\subsection{Polarization Image Formation (HDR)}
When light reflects off a non-metallic object, it becomes partially polarized. This polarized light can be captured by a polarization camera. In the polarization camera \cite{Polcam}, the image irradiance $I_0$ is first filtered by the on-chip directional micro-polarizers. Then the camera converts the light signal, using the camera response function (CRF), into four digital images $L_1$ to $L_4$. The effect of a polarizer on image irradiance can be written as \cite{Goldstein}:
\begin{equation}
I_i = 0.5 \cdot I_0 \Big(1 + \rho\cos(2\theta - 2\alpha_i)\Big)
\label{eqn:image_irradiance}
\end{equation}
\noindent where $\rho$ is the degree of polarization (DoP) in [0,1], $\theta$ is the angle of polarization (AoP) in [0$^{\circ}$, 180$^{\circ}$], $\alpha_i$ = 0$^{\circ}$, 45$^{\circ}$, 90$^{\circ}$, 135$^{\circ}$ denotes the angle of the polarizers, and $i=$ 1, 2, 3, 4 denotes the index of the four polarizers. Substituting the polarizer angles into Eq. \ref{eqn:image_irradiance}, we get the filtered images:
\begin{equation}
\begin{gathered}
I_1, I_3 = 0.5 \cdot I_0(1 \pm \rho\cos2\theta)\\
I_2, I_4 = 0.5 \cdot I_0(1 \pm \rho\sin2\theta)\\
\label{eqn:polarizer_irradiance}
\end{gathered}
\end{equation}
\vspace{-0.6cm}

\noindent Then $\rho$ and $\theta$ can be computed from the Stokes by \cite{Goldstein}:
\begin{equation}
\rho = \frac{\sqrt{S_1^2 + S_2^2}}{S_0}, \quad  \theta = 0.5 \cdot \tan^{-1}\Big(\frac{S_2}{S_1}\Big)
\label{eqn:dop_aop}
\end{equation}

In general, the relation between image irradiance $I_i$ and pixel value $L_i$ at exposure time $t_0$ can be written as \cite{Debevec}:
\begin{equation}
L_i = f(I_{i}t_0)
\label{eqn:irradiance_pixel}
\end{equation}

\noindent where $f$ is the CRF. By substituting Eq. \ref{eqn:polarizer_irradiance} into Eq. \ref{eqn:irradiance_pixel}, and applying the reciprocity relation in \cite{Debevec}, we obtain:
\begin{equation}
\begin{gathered}
t_1, t_3 = 0.5 \cdot t_0 (1 \pm \rho\cos2\theta)\\
t_2, t_4 = 0.5 \cdot t_0 (1 \pm \rho\sin2\theta)\\
\label{eqn:HDR}
\end{gathered}
\end{equation}

\vspace{-0.6cm}
Eq. \ref{eqn:HDR} provides the polarization image formation model in the case of HDR reconstruction. When the incoming light is not entirely unpolarized ($\rho \neq$ 0), the four pixels within one calculation unit of the polarization camera experience different exposure times, effectively creating the condition for multiple exposures. In the extreme case, when $\rho \approx$ 1 one can expect a large difference in exposure between the four pixels.

Different from a conventional camera capturing multiple exposures, the variation in exposure time in a polarization camera is pixel specific as both $\rho$ and $\theta$ vary from pixel to pixel. This is because the pixels in different calculation units of a polarization camera correspond to different points in space, which in general differ in terms of their light polarization, just as in color or in intensity. A comparative study on the increase in dynamic range of a conventional camera versus a polarization camera is performed by \cite{xuesong}, in which the polarization camera shows superiority for HDR reconstruction.

\begin{figure*}[t!]
\centering
\begin{subfigure}{.15\textwidth}
  \centering
  \includegraphics[width=0.95\linewidth]{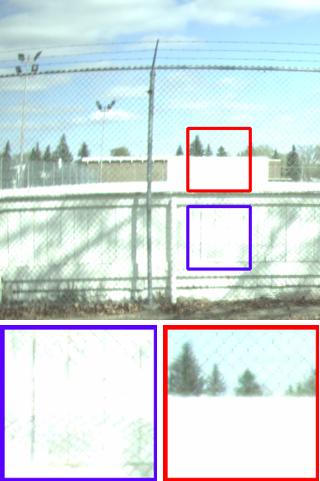}
  \caption{Input LDR}
  \label{fig:input_ldr}
\end{subfigure}%
\begin{subfigure}{.15\textwidth}
  \centering
  \includegraphics[width=0.95\linewidth]{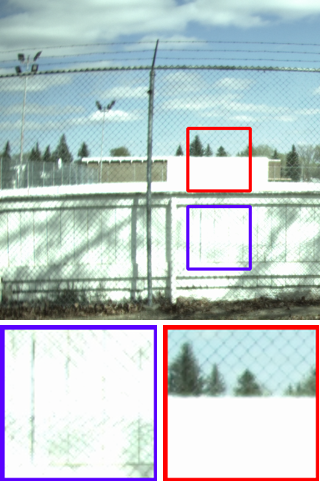}
  \caption{$I_{Deb}$}
  \label{fig:input_I}
\end{subfigure}%
\begin{subfigure}{.15\textwidth}
  \centering
  \includegraphics[width=0.95\linewidth]{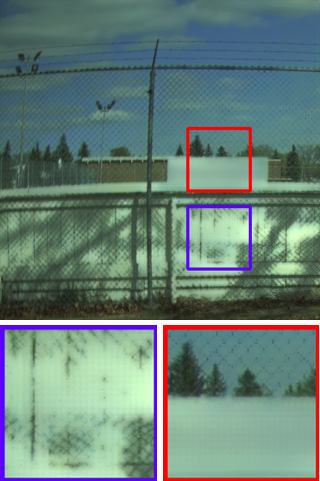}
  \caption{Ours w/o $I_{Deb}$}
  \label{fig:1PHDRNet}
\end{subfigure}%
\begin{subfigure}{.15\textwidth}
  \centering
  \includegraphics[width=0.95\linewidth]{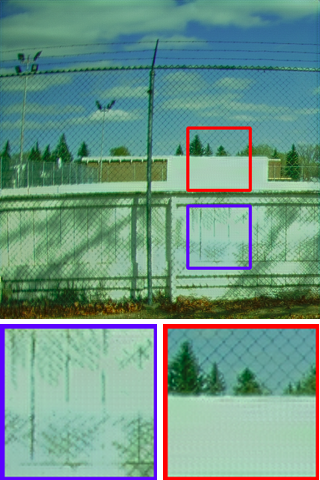}
  \caption{Ours}
  \label{fig:4PHDRNet}
\end{subfigure}%
\begin{subfigure}{.15\textwidth}
  \centering
  \includegraphics[width=0.95\linewidth]{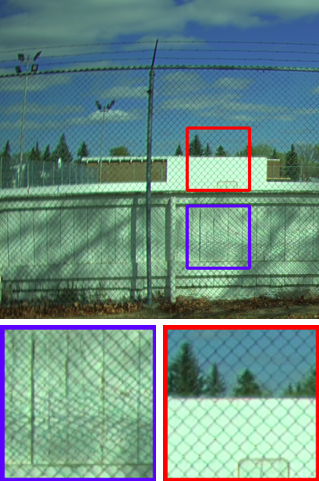}
  \caption{Ground truth}
  \label{fig:ablation_ground_truth}
\end{subfigure}%
\caption{Qualitative results of model variants, where image details are better restored from left to right.}
\vspace{-0.5\baselineskip}
\label{fig:ablation_model}
\end{figure*}

\vspace{-0.4cm}
\subsection{Framework Overview}
Fig.~\ref{fig:PFHDRNet} illustrates our framework. In order to reduce computation parameters and improve detail reconstruction, we propose a pre-process step to fuse the four LDR polarization images before feeding it to the network. Based on Eq. \ref{eqn:HDR}, the fusion step can be implemented by a pixel-weighting function which assigns weights to the four pixels in one calculation unit depending on how well-exposed the pixels are as was proposed by Debevec in \cite{Debevec} and adopted in \cite{xuesong}. The resulting image $I_{Deb}$ can be computed by:
\begin{equation}
I_{Deb} =\frac{\sum_{i=1}^{2}W\Big(L_{i}+L_{i+2}\Big)\Big(g\big(L_{i}\big)+g\big(L_{i+2}\big)\Big)}{\sum_{i=1}^{2}W\Big(L_{i}+L_{i+2}\Big)t_0} \label{eqn:pixel_weight_hdr}
\end{equation}

\noindent where $W$ is the Gaussian weighted function ($\sigma$ = 0.2 in our study), and $g$ is the inverse-CRF. Intuitively, Eq. \ref{eqn:pixel_weight_hdr} transforms the input to the HDR space, which conveys more information as $I_{Deb}$ is in floating point and spans a wider dynamic range than $I_i$.

We adopt an 18 level autoencoder architecture for our network \cite{Endo}. The architecture generates multiexposure images, and then merges the images to output an HDR image. In addition, we introduce the following improvements: 1) We replace the deconvolution operation with a bilinear interpolation operation. This reduces the visible tiling artefacts when the input image has large saturated regions \cite{checkerboard_artefacts}. 2) We add the SSIM loss to the loss function. This additional loss term improves the perceptual quality of the synthesized content \cite{ssim}. Our overall loss function is:
\begin{equation}
\mathcal{L}(W,b) = \alpha \cdot \mathbb{E}[|H-\hat{H}|] + \beta \cdot SSIM(H,\hat{H})
\label{eqn:loss}
\end{equation}

\noindent where $H$ and $\hat{H}$ denote the ground truth and prediction image, respectively. The first term in Eq. \ref{eqn:loss} corresponds to the mean absolute error (MAE) loss which measures the pixel-level loss between the prediction and ground truth images. By minimizing the MAE loss, details can be better recovered. The second term in Eq. \ref{eqn:loss} corresponds to the SSIM loss, which measures the perceptual loss between the prediction and ground truth images. By maximizing the SSIM loss, edges and contrast details can be better preserved. We empirically set $\alpha = 1$, and $\beta = 0.1$ in our experiments.

\vspace{-0.3cm}
\section{Experiments} \label{sec:Experiments}
\subsection{Experiment Setups}
\noindent \textbf{Dataset.} Since there is no public dataset available for training and testing HDR techniques with polarized images, we collected an outdoor dataset, under sunlight conditions, named EdPolCommunity. Images are captured using the FLIR BFS-U3-51S5p polarization camera. The dataset consists of colored images of four polarizer orientations capturing each scene with 17 exposure times ($t_0$ = 0.03, 0.045, 0.068, 0.101, 0.152, 0.228, 0.342, 0.513, 0.769, 1.153, 1.73, 2.595, 3.592, 5.839, 8.758, 13.137, 19.705 ms). The resolution of each polarization image is 1024x1224. Then four 512x512 patches are cropped from each image for better visibility. In total, our EdPolCommunity dataset has 41,616 LDR images and 612 HDR image as the ground truth to train and test the network. The ground truth HDR images are created by combining the four polarization images using Eq. \ref{eqn:pixel_weight_hdr}, and then merging the bracketed images using the method in \cite{Mertens} and the inverse-CRF. The ground truth LDR images, useful for visualization, are generated with Reinhard \cite{Reinhard} tone mapping.

\smallskip
\noindent \textbf{Implementation Details.} The framework is implemented using Chainer where the up- and down-exposure models are trained on NVIDIA GTX 1080 TI. We trained using an Adam optimizer \cite{Adam} and set the batch size, learning rate and momentum as 1, 0.0002 and 0.5, respectively. We initialized all weights with zero mean Gaussian noise ($\sigma$ = 0.02). 

\smallskip
\noindent \textbf{Evaluation Metrics.} We evaluate the accuracy of HDR reconstruction with the HDR-VDP2 \cite{hdrvdp2} metric. We normalize the predicted and reference ground truth HDR images \cite{Marnerides}. The ground truth images capture wide dynamic range scene contents. We also use the PSNR, SSIM \cite{ssim} and FSIM \cite{fsim} metrics to evaluate the image quality on the HDR tone mapped images.

\begin{table}[t!]
\caption{Quantitative results of model variants.}
\vspace{-1.5\baselineskip}
\begin{center}
\label{tab:ablation_model}
\begin{tabular}{c|c|c|c|c}
\hline
& \small{PSNR} & \small{SSIM} & \small{FSIM} & \small{HDR-VDP2}\\ 
\hline
\small{Input LDR} & 11.76 & 0.59 & 0.87 & 44.85\\
\small{$I_{Deb}$ \cite{xuesong}} & 15.42 & 0.67 & 0.89 & 45.72\\
\hline
\small{Ours w/o $I_{Deb}$} & 20.61 & 0.80 & 0.90 & 49.56\\ 
\small{\textbf{Ours}} & \textbf{25.35} & \textbf{0.91} & \textbf{0.94} & \textbf{55.26}\\ 
\hline
\end{tabular}
\end{center}
\vspace{-1.5\baselineskip}
\end{table}

\vspace{-0.37cm}
\subsection{Ablation Studies}
\subsubsection{Study on the Model Architecture}
We validate the importance of different individual components in our framework. Fig.~\ref{fig:ablation_model} shows the result for the ablation study. We can observe that the case without $I_{Deb}$ reconstructs fewer details compared with the case with $I_{Deb}$. It is because the case without $I_{Deb}$ uses a single image as input which conveys less information, and thus tend to neglect textures and local contrasts. On the other hand, $I_{Deb}$ constructs an intermediate HDR image, from four polarizer orientation images, which conveys an information rich input. These four images effectively correspond to four different exposure times on a per-pixel basis, and can thus better reveal scene details and color contrast. As shown by the quantitative results in Table~\ref{tab:ablation_model}, our framework (last row) achieves the best results. In addition, the deconvolution interpolation operation produces clearer and smoother images.

We also study the benefit of the fusion mechanism by comparing two different input images, namely $I_{Deb}$ \cite{Debevec} and input LDR. As displayed in Fig.~\ref{fig:input_ldr} and \ref{fig:input_I}, $I_{Deb}$ overall provides more details than input LDR, and thus is a more informative and better input. Quantitative comparisons of the results are shown in Table~\ref{tab:ablation_model}.
\vspace{-0.3cm}
\subsubsection{Study on the Loss Function} \label{sssec:ablation_loss}

\begin{figure}
\centering
\begin{subfigure}{.23\textwidth}
  \centering
  \includegraphics[width=0.95\linewidth]{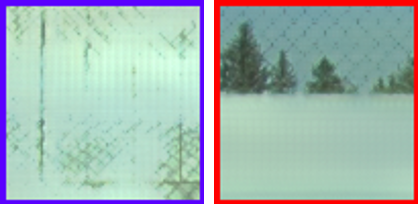}
  \caption{$\ell_1$ loss}
  \label{fig:L1_loss}
\end{subfigure}%
\begin{subfigure}{.23\textwidth}
  \centering
  \includegraphics[width=0.95\linewidth]{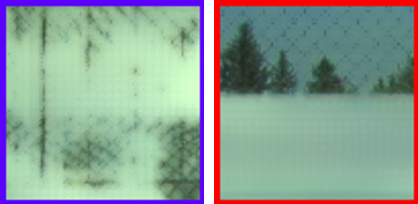}
  \caption{$\ell_1 + SSIM$ loss}
  \label{fig:L1SSIM_loss}
\end{subfigure}%
\caption{Qualitative results of different loss functions, where our choice of $\ell_1 + SSIM$ loss restores more details.}
\label{fig:ablation_loss}
\end{figure}

\begin{table}[t!]
\caption{Quantitative results of different loss functions}
\vspace{-1.5\baselineskip}
\begin{center}
\label{tab:ablation_loss}
\begin{tabular}{c|c|c|c|c}
\hline
& \small{PSNR} & \small{SSIM} & \small{FSIM} & \small{HDR-VDP2}\\ 
\hline
\small{$\ell_1$} & 17.64 & 0.78 & 0.91 & 48.48\\ 
\small{\pmb{$\ell_1$} \textbf{+ SSIM}} & \textbf{20.61} & \textbf{0.80} & \textbf{0.90} & \textbf{49.56}\\
\hline
\end{tabular}
\end{center}
\vspace{-1.5\baselineskip}
\end{table}

In this experiment, we compare the performances of our method with different loss functions. For simplicity, we use the model \textit{ours w/o $I_{Deb}$} to analyze the effect of each loss function. Qualitative results are illustrated in Fig.~\ref{fig:ablation_loss} where $\ell_1 + SSIM$ loss is better at preserving details. This is also reflected by the quantitative results shown in Table~\ref{tab:ablation_loss}. Therefore, we train our model using $\ell_1 + SSIM$ loss.

\vspace{-0.3cm}
\subsection{Comparison with State-of-the-art Methods}
We compare the proposed method with four state-of-the-art learning based methods: HDRCNN \cite{Eilertsen}, ENet \cite{Marnerides}, DrTMO \cite{Endo} and DRCP \cite{Liu}. For all the methods, we use source code and pre-trained model provided by the authors, with a single polarization image as input. We also compare our method with PHDR \cite{xuesong}, a non-learning method developed directly for polarization based HDR recovery. 

\begin{figure}
\centering
\begin{subfigure}{.23\textwidth}
  \centering
  \includegraphics[width=0.95\linewidth]{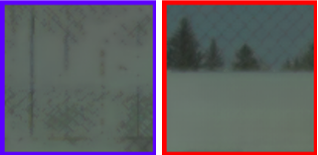}
  \caption{HDRCNN}
  \label{fig:HDRCNN}
\end{subfigure}%
\begin{subfigure}{.23\textwidth}
  \centering
  \includegraphics[width=0.95\linewidth]{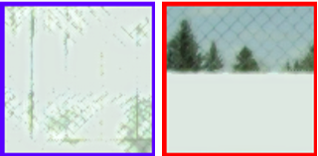}
  \caption{ENet}
  \label{fig:ENET}
\end{subfigure}%
\vskip 0pt
\begin{subfigure}{.23\textwidth}
  \centering
  \includegraphics[width=0.95\linewidth]{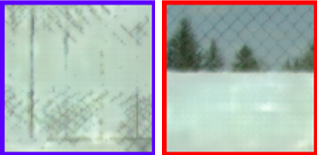}
  \caption{DrTMO}
  \label{fig:DRTMO}
\end{subfigure}%
\begin{subfigure}{.23\textwidth}
  \centering
  \includegraphics[width=0.95\linewidth]{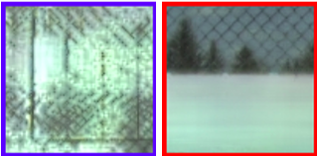}
  \caption{DRCP}
  \label{fig:DRCP}
\end{subfigure}%
\vskip 0pt
\begin{subfigure}{.23\textwidth}
  \centering
  \includegraphics[width=0.95\linewidth]{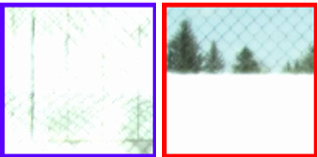}
  \caption{PHDR}
  \label{fig:soa_PHDR}
\end{subfigure}%
\begin{subfigure}{.23\textwidth}
  \centering
  \includegraphics[width=0.95\linewidth]{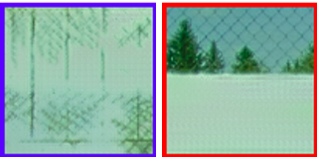}
  \caption{Ours}
  \label{fig:soa_PAFHDRNet}
\end{subfigure}%
\caption{Qualitative comparative results with state-of-the-art methods. Our framework recovers the most details, and is close to the ground truth image.}
\vspace{-1\baselineskip}
\label{fig:soa_results}
\end{figure}

Qualitative results are shown in Fig.~\ref{fig:soa_results}. HDRCNN \cite{Eilertsen} results tend to be dim, and the network is unable to restore details in the saturated regions. ENet \cite{Marnerides} generates overly-bright and smooth results, as it over-enhances the extracted illumination features. It also fails to recover details which reside in the overexposed regions. The results of DrTMO \cite{Endo} suffer from blocking artefacts and can not preserve details in the saturated areas. DRCP \cite{Liu} shares similar limitations and the results lack color consistency, as in some cases the generated colors are unnatural with artefacts. PHDR \cite{xuesong} results tend to be bright, and the method cannot recover the information in the saturated regions. Since our method fuses four polarization images captured in a snapshot, where the images also correspond to those captured under four different exposure times, the given input is able to utilize information in the unsaturated pixels from one or more of the images to reveal details. As a result, the polarized input images collectively convey richer details compared to an image taken by a conventional camera. This fusion mechanism helps to reconstruct details, and outputs visually pleasing textures. 

In addition to visual evaluation, the quantitative results are summarized in Table~\ref{tab:soa_results}. It shows that our method performs favorably compared to state-of-the-art methods under various evaluation metrics. 

\begin{table}[t!]
\caption{Quantitative comparative results with state-of-the-art methods.}
\vspace{-1.5\baselineskip}
\begin{center}
\label{tab:soa_results}
\begin{tabular}{c|c|c|c|c}
\hline
\small{Methods} & \small{PSNR} & \small{SSIM} & \small{FSIM} & \small{HDR-VDP2}\\ 
\hline
\small{HDRCNN \cite{Eilertsen}} & 14.18 & 0.40 & 0.71 & 47.49\\ 
\small{ENet \cite{Marnerides}} & 15.37 & 0.67 & 0.88 & 47.24\\
\small{DrTMO \cite{Endo}} & 16.16 & 0.64 & 0.89 & 47.70\\ 
\small{DRCP \cite{Liu}} & 17.24 & 0.70 & 0.90 & 51.46\\
\small{PHDR \cite{xuesong}} & 15.42 & 0.67 & 0.89 & 45.72\\
\hline
\small{\textbf{Ours}} & \textbf{25.35} & \textbf{0.91} & \textbf{0.94} & \textbf{55.26}\\
\hline
\end{tabular}
\end{center}
\vspace{-1.2\baselineskip}
\end{table}

\vspace{-0.5cm}
\section{Conclusion}
This paper presented a deep framework to address the snapshot HDR reconstruction problem using polarization images. It effectively utilizes the benefits of HDR imaging hardware and deep learning to restore details for snapshot HDR image reconstruction. Our key insight is to leverage the domain knowledge of polarization for designing the framework. We show that the our framework demonstrated better quantitative and qualitative performances over state-of-the-art methods. In addition, we introduced a new polarized HDR dataset for training and testing HDR techniques, which will be made publicly available with the publication of this paper.

To follow up on this study, we plan to explore the use of DoP and AoP images to further the image dynamic range and detail recovery. It would also be interesting to investigate the transferability of this approach to underexposed polarized datasets for HDR reconstruction.

\vspace{-0.2cm}
{\small
\bibliographystyle{IEEEbib}
\bibliography{mybib}

\begin{thebibliography}{10}

\bibitem{Lin_objectDetect}
T.~{Lin}, P.~{Dollár}, R.~{Girshick}, K.~{He}, B.~{Hariharan}, and
  S.~{Belongie},
\newblock ``Feature pyramid networks for object detection,''
\newblock in {\em 2017 IEEE Conference on Computer Vision and Pattern
  Recognition (CVPR)}, 2017, pp. 936--944.

\bibitem{Song_visualTrack}
Y.~{Song}, C.~{Ma}, L.~{Gong}, J.~{Zhang}, R.~W.~H. {Lau}, and M.~{Yang},
\newblock ``Crest: Convolutional residual learning for visual tracking,''
\newblock in {\em 2017 IEEE International Conference on Computer Vision
  (ICCV)}, 2017, pp. 2574--2583.

\bibitem{Debevec}
Paul~E. Debevec and Jitendra Malik,
\newblock ``Recovering high dynamic range radiance maps from photographs,''
\newblock in {\em Proceedings of the 24th Annual Conference on Computer
  Graphics and Interactive Techniques}, USA, 1997, SIGGRAPH '97, p. 369–378,
  ACM Press/Addison-Wesley Publishing Co.

\bibitem{Mann}
S.~Mann and R.~W. Picard,
\newblock ``On being 'undigital' with digital cameras: Extending dynamic range
  by combining differently exposed pictures,''
\newblock in {\em PROCEEDINGS OF IST}, 1995, pp. 442--448.

\bibitem{Yan_AttnHDR}
Q.~{Yan}, D.~{Gong}, Q.~{Shi}, A.~{van den Hengel}, C.~{Shen}, I.~{Reid}, and
  Y.~{Zhang},
\newblock ``Attention-guided network for ghost-free high dynamic range
  imaging,''
\newblock in {\em 2019 IEEE/CVF Conference on Computer Vision and Pattern
  Recognition (CVPR)}, 2019, pp. 1751--1760.

\bibitem{Lee_ExpBlendHDR}
S.~{Lee}, H.~{Chung}, and N.~I. {Cho},
\newblock ``Exposure-structure blending network for high dynamic range imaging
  of dynamic scenes,''
\newblock {\em IEEE Access}, vol. 8, pp. 117428--117438, 2020.

\bibitem{Akyuz}
A.~Aky{\"u}z, R.~Fleming, Bernhard~E. Riecke, E.~Reinhard, and H.~B{\"u}lthoff,
\newblock ``Do hdr displays support ldr content?: a psychophysical
  evaluation,''
\newblock {\em ACM Trans. Graph.}, vol. 26, pp. 38, 2007.

\bibitem{Masia}
Belen Masia, Sandra Agustin, Roland~W. Fleming, Olga Sorkine, and Diego
  Gutierrez,
\newblock ``Evaluation of reverse tone mapping through varying exposure
  conditions,''
\newblock in {\em ACM SIGGRAPH Asia 2009 Papers}, New York, NY, USA, 2009,
  SIGGRAPH Asia '09, Association for Computing Machinery.

\bibitem{Eilertsen}
Gabriel Eilertsen, Joel Kronander, Gyorgy Denes, Rafał~K. Mantiuk, and Jonas
  Unger,
\newblock ``Hdr image reconstruction from a single exposure using deep cnns,''
\newblock {\em ACM Transactions on Graphics}, vol. 36, no. 6, pp. 1–15, Nov
  2017.

\bibitem{Marnerides}
Demetris Marnerides, Thomas Bashford-Rogers, Jonathan Hatchett, and Kurt
  Debattista,
\newblock ``Expandnet: A deep convolutional neural network for high dynamic
  range expansion from low dynamic range content,''
\newblock 2019.

\bibitem{Liu}
Y.~L. {Liu}, W.~S. {Lai}, Y.~S. {Chen}, Y.~L. {Kao}, M.~H. {Yang}, Y.~Y.
  {Chuang}, and J.~B. {Huang},
\newblock ``Single-image hdr reconstruction by learning to reverse the camera
  pipeline,''
\newblock in {\em 2020 IEEE/CVF Conference on Computer Vision and Pattern
  Recognition (CVPR)}, 2020, pp. 1648--1657.

\bibitem{Endo}
Yuki Endo, Yoshihiro Kanamori, and Jun Mitani,
\newblock ``Deep reverse tone mapping,''
\newblock {\em ACM Transactions on Graphics (Proc. of SIGGRAPH ASIA 2017)},
  vol. 36, no. 6, Nov. 2017.

\bibitem{Nayar}
S.~K. {Nayar} and T.~{Mitsunaga},
\newblock ``High dynamic range imaging: spatially varying pixel exposures,''
\newblock in {\em Proceedings IEEE Conference on Computer Vision and Pattern
  Recognition. CVPR 2000 (Cat. No.PR00662)}, 2000, vol.~1, pp. 472--479 vol.1.

\bibitem{Suda}
Takeru Suda, Masayuki Tanaka, Yusuke Monno, and Masatoshi Okutomi,
\newblock ``Deep snapshot hdr imaging using multi-exposure color filter
  array,'' 2020.

\bibitem{xuesong}
X.~{Wu}, H.~{Zhang}, X.~{Hu}, M.~{Shakeri}, C.~{Fan}, and J.~{Ting},
\newblock ``Hdr reconstruction based on the polarization camera,''
\newblock {\em IEEE Robotics and Automation Letters}, vol. 5, no. 4, pp.
  5113--5119, 2020.

\bibitem{Polcam}
``Imx250 cmos sensor,''
  \url{https://www.sony-semicon.co.jp/e/products/IS/industry/technology/polarization.html}.

\bibitem{Goldstein}
D.H. {Goldstein},
\newblock {\em Polarized light},
\newblock 2017.

\bibitem{checkerboard_artefacts}
Augustus Odena, Vincent Dumoulin, and Chris Olah,
\newblock ``Deconvolution and checkerboard artifacts,''
\newblock {\em Distill}, 2016.

\bibitem{ssim}
{Zhou Wang}, A.~C. {Bovik}, H.~R. {Sheikh}, and E.~P. {Simoncelli},
\newblock ``Image quality assessment: from error visibility to structural
  similarity,''
\newblock {\em IEEE Transactions on Image Processing}, vol. 13, no. 4, pp.
  600--612, 2004.

\bibitem{Mertens}
T.~{Mertens}, J.~{Kautz}, and F.~{Van Reeth},
\newblock ``Exposure fusion,''
\newblock in {\em 15th Pacific Conference on Computer Graphics and Applications
  (PG'07)}, 2007, pp. 382--390.

\bibitem{Reinhard}
Erik Reinhard, Greg Ward, Sumanta Pattanaik, Paul Debevec, and Wolfgang
  Heidrich,
\newblock {\em High dynamic range imaging : acquisition, display, and
  image-based lighting},
\newblock 01 2010.

\bibitem{Adam}
Diederik~P. Kingma and Jimmy Ba,
\newblock ``Adam: A method for stochastic optimization,'' 2017.

\bibitem{hdrvdp2}
Rafat Mantiuk, Kil~Joong Kim, Allan~G. Rempel, and Wolfgang Heidrich,
\newblock ``Hdr-vdp-2: A calibrated visual metric for visibility and quality
  predictions in all luminance conditions,''
\newblock {\em ACM Trans. Graph.}, vol. 30, no. 4, July 2011.

\bibitem{fsim}
L.~{Zhang}, L.~{Zhang}, X.~{Mou}, and D.~{Zhang},
\newblock ``Fsim: A feature similarity index for image quality assessment,''
\newblock {\em IEEE Transactions on Image Processing}, vol. 20, no. 8, pp.
  2378--2386, 2011.

\end{thebibliography}
}

\end{document}